# A Quantitative Overview of Biophysical Forces Governing Neural Function

Jerel Mueller[1,2] and William J. Tyler[1,2*]

[1] Virginia Tech Carilion Research Institute
[2] Virginia Tech - Wake Forest School of Biomedical Engineering and Sciences

The Hodgkin-Huxley (HH) model is the currently accepted formalism of neuronal excitability. However, the HH model does not capture a number of biophysical behaviors associated with action potentials or propagating nerve impulses. Physical mechanisms underlying these processes, such as reversible heat transfer and axonal swelling have been separately investigated and compartmentally modeled to indicate the nervous system is not purely electrical or biochemical. Rather, mechanical forces and principles of thermodynamics also govern neuronal excitability and signaling. To advance our understanding of neural function and dysfunction, compartmentalized analyses of electrical, chemical, and mechanical processes need to revaluated and integrated into more comprehensive theories. The present quantitative perspective is intended to broaden the awareness of known biophysical phenomena, which are often overlooked in neuroscience. By starting to consider the collective influence of the biophysical forces influencing neural function, new paradigms can be applied to the characterization and manipulation of nervous systems.

## Introduction

The physiological functions of nervous systems are primarily regarded as being driven by electrical and chemical driving forces. For instance, we have an intimate portrait of how electrical signaling along axonal fibers is converted to chemical signaling at the synapse between neurons. It is not well understood however how other forces, such as mechanical ones, impart actions upon neural function. Thus, to advance our understanding of how nervous systems operate it is important to develop comprehensive models where electrical, chemical, and mechanical energies are not compartmentalized from one another, but rather cooperate in a synergistic manner to govern neuronal excitability and signaling.

The consideration of mechanical forces on neuronal function has been of great interest over the past several decades [1] and continues to grow while gaining support through its incorporation in system characterization and manipulation [2]. Outside of neuroscience, there exist well-established models describing the interplay between mechanical and electro-chemical signaling [3-7]. In fact the importance of mechanical cues on physiology is highlighted in several organ systems. For instance the influences of mechanical forces on heart function is broadly implicated, and stretch-activated channels (SACs) are known to play a role in cardiac pacing [8-12]. The colon [13], bladder [14], and muscles [15] also generate ordered signals through SAC and mechanosensitive channel (MSC) activity. In sensory neuroscience it is broadly recognized that MSCs are involved in signal transduction processes in hair cells for hearing [16] and free nerve endings for touch [17].

Numerous mechanical events have been observed and associated with neuronal activity. However, only recently has technology emerged capable of observing and modulating mechanical events in nervous tissues. The quantitative perspective provided by the present article includes an overview of biophysical events known to occur in the nervous system. We also highlight quantitative formulations of the influence mechanical forces have on neuronal function. From these formulations, models and technology for mechanical manipulation and interfacing with the nervous system can be further developed. By starting to consider the interplay between electrical, chemical, and mechanical energy, rather than separately compartmentalizing them, new paradigms for understanding and studying the biophysics of neural systems should advance.

## Electro-mechanical coupling and deformation forces

The Hodgkin-Huxley (HH) model is a bioelectric description of neuronal excitability based on conductance of ion-selective channels and a membrane capacitor, and is the currently accepted model for describing the action potential [18]. However, there are a number of observations related to the action potential that are not electrical or electro-chemical in nature. Several studies have shown the geometric dimensions of nerve fibers change in phase with the action potential propagation, exerting forces normal to the membrane surface [1, 19-23]. Additionally, there is a reversible change in heat generation during action potential propagation, where heat released during the first phase of the action potential is compensated by heat uptake in the second phase [23-27]. The HH model however is based on irreversible processes and does not include thermodynamic variables required to sufficiently explain all the physically observed features of a nerve impulse. Despite this shortcoming the equivalent RC circuit formalism of the HH model [18] has, no doubt, acquired global support through a bewildering number of independent observations over the past 60 years. While separate models accounting for the other non-electrical behaviors observed during the action potential have been proposed, there is not a broadly accepted model unifying electrical, chemical, and mechanical descriptions of the neuronal action potential. Regardless, the coupling of mechanical and electrical energy has seen considerable research and development (for example, piezoelectricity) and its consideration as applied to the nervous system is briefly highlighted below.

* Correspondence to WJT wtyler@tylerlab.com

### The flexoelectric effect

The flexoelectric effect is a liquid crystal analogue to the piezoelectric effect in solid crystals. Flexoelectricity refers specifically to the curvature dependent polarization of the membrane [28]. As opposed to area stretching, thickness compression, and shear deformation in solid crystals, the flexoelectric effect includes the deformation of membrane curvature. This effect is manifested in liquid crystalline membrane structures, as a curvature of membrane surface leads to a splay of lipids and proteins. The molecules would otherwise be oriented parallel to each other in the normal flat state of the local membrane. Similar to piezoelectricity of solids, flexoelectricity is also manifested as a direct and a converse effect, featuring electric field induced curvature. The flexoelectric effect provides a basic mechanoelectric mechanism enabling nanometer-thick biomembranes to exchange responsiveness between electrical and mechanical stimuli. Consideration that cellular membranes possess mechanoelectric properties has raised concerns regarding the possible origin of inductance in early circuit models of the neuronal membrane and giant squid axons [29].

Experimentally, the generation of alternating currents by membranes subjected to oscillating gradients of hydrostatic pressure was observed in the early 1970's [30]. This observed vibratory response was assumed to be due to changes in membrane area and thus capacitance, though a detailed explanation of the mechanisms with regard to transmembrane potential was not offered at the time. The oscillations of membrane curvature in these experiments can be credited as a displacement current due to the oscillating reversal of flexoelectric polarization of the curved membrane [28]. According to the Helmholtz equation, an electric potential difference appears across a polarized surface. For a membrane curvature that oscillates in time, and assuming spherical curvature for simplicity, the flexopolarization leads to a transmembrane AC voltage difference with first harmonic amplitude described by Equation 1:

$$U_\omega = \frac{f}{\varepsilon_0} 2 c_m \quad (1)$$

where $f$ is the flexoelectric coefficient, measured in coulombs, $\varepsilon_0$ is the absolute dielectric permittivity of free space, and $c_m$ is the maximal curvature. Similarly, a displacement current due to oscillating flexopolarization can also be calculated by considering a membrane capacitance described by Equation 2:

$$C_0 = \frac{\varepsilon_0 S_0}{d} \quad (2)$$

where $S_0$ is the flat membrane area and $d$ is the capacitive thickness of the membrane. The first harmonic amplitude of the membrane flexoelectric current can then be described by Equation 3:

$$I_\omega = f \frac{C_0}{\varepsilon_0} 2 c_m \omega \quad (3)$$

where $\omega$ is the angular frequency of oscillations. Thus, the current through or potential across the membrane can be determined from the associated flexoelectric coefficient of the membrane and the radius of curvature. Flexoelectricity (current generation from bending) and converse flexoelectricity have been demonstrated in lipid bilayers and cell membranes [28, 31]. Flexoelectricity provides a linear relationship between membrane curvature and transmembrane voltage and is likely involved in mechanosensitivity and mechanotransduction in biological systems [31]. The direct and converse flexoelectric effects have been used to describe the transformation of mechanical into electrical energy by stereocila and the electromotility of outer hair cell membranes for hearing [28, 31]. Many membrane functions involve the manipulation of membrane curvature (for example, exocytosis, endocytosis, and cell migration) and the prospects that flexoelectricity is intricately involved in these processes thereby relating membrane mechanics and electrodynamics is likely.

### Voltage-induced changes in membrane tension

Mechanical equilibrium in membranes requires that the cellular radius depend on membrane tension in order to maintain a constant pressure across the membrane, as related by the Young-Laplace equation:

$$\Delta P = \gamma \left( \frac{1}{R_1} + \frac{1}{R_2} \right) \quad (4)$$

where $\Delta P$ is the pressure difference across the membrane, $\gamma$ is the membrane tension, and $R_1$ and $R_2$ are the principal radii of curvature. Considering now electric field mediated effects on the membrane, the relation between membrane tension and an applied electrostatic potential is given by the Young-Lippmann equation:

$$\gamma = \gamma^0 - \frac{CV^2}{2} \quad (5)$$

where $\gamma$ is the total chemical and electrical surface tension, $\gamma^0$ is the surface tension at zero electric field, $C$ is the capacitance of the interface, and $V$ is the applied voltage. Electrowetting on dielectrics is one application concerned with membrane tension as related to an applied voltage [32-34]. In electrowetting, a thin insulating layer (analogous to the cell membrane) is used to separate conductive liquid (extracellular environment) from metallic electrodes with an applied voltage (intracellular environment) to avoid electrolysis.

In the case of a phospholipid bilayer cellular membrane, differences in tension between the two interfaces will create changes in curvature, referred to earlier as converse flexoelectricity. Thus, modulation of membrane tension by transmembrane voltage in a neuron should cause movement of the membrane with magnitude and polarity governed by the cell membrane stiffness and surface potentials in order to maintain pressure across the membrane. Such an effect has been observed in real-time using atomic force microscopy (AFM) and voltage clamped HEK293 cells [35]. In these studies Zhang et al. (2001) observed that depolarization caused an outward movement of the membrane, with amplitude proportional to voltage. Additionally application of the Young-Lippmann equation (Equation 5) to both interfaces of the lipid bilayer yielded a mathematical model able to predict the membrane tension over a range of surface potentials [35]. The sum of the two interface tensions yields the total tension in the membrane ($T_t$) as described by Equation 6:

$$T_t = \frac{\sqrt{(2k_BT)^3 \epsilon_w \epsilon_0}}{ze_0} \left( \sqrt{n_{ex}} \left[ \sinh^{-1}\left\{ \frac{\sigma_{ex} - C_m V}{2\sqrt{n_{ex}\epsilon_w\epsilon_0 2k_BT}} \right\} \right]^2 + \sqrt{n_{in}} \left[ \sinh^{-1}\left\{ \frac{\sigma_{in} + C_m V}{2\sqrt{n_{in}\epsilon_w\epsilon_0 2k_BT}} \right\} \right]^2 \right) + T_m \quad (6)$$

where $k_B$ is Boltzmann's constant, $T$ is the absolute temperature, $\epsilon_0$ the permittivity of free space, $\epsilon_w$ the relative permittivity of water, $z$ ion valence, $e_0$ electronic charge, $n$ ionic strength of the solution, $\sigma$ structural charge density at the interface, $C_m$ specific capacitance of the membrane, $V$ voltage, $T_m$ the voltage independent portion of membrane tension, and the subscripts $ex$ and $in$ refer to the external and internal membrane interfaces respectively. Thus, the tension in the membrane is related to the voltage and ionic charges across the membrane. Consequently, the change in voltage with a nerve impulse is associated with a change in membrane tension, which will result in an alteration of cell radius to keep pressure constant across the membrane. This offers a mechanism and quantitative description for the observed change in the diameter of nerve fibers during the action potential, as opposed to alternative hypothesized mechanisms such as cell swelling due to water transport [20].

## Opto-electric and electro-mechanical coupling

Our early understanding of the phenomena of electrical coupling with the mechanical modification of the neuronal membrane has already begun to yield innovative methods and technologies for interfacing to the nervous system. The modulation of refractive index or thickness of the cell due to transmembrane potential dependent deformations has allowed label free imaging of the membrane potential without the need of organic dyes or optogenetic probes which themselves likely alter membrane dynamics [36]. By measuring milliradian scale phase shifts in the transmitted light, changes in the membrane potential of individual mammalian cells have been detected using low coherence interferometric microscopy without the use of exogenous labels [36]. Using this technique, it was also demonstrated that propagation of electrical stimuli in gap junction-coupled cells could be monitored using wide-field imaging. This technique offers the advantages of simple sample preparation, low phototoxicity, and no need for photobleaching. Previous successes in label-free imaging of electrical activity has been possible in invertebrate nerves and neurons, as mammalian cells are smaller, optically transparent, and scatter light significantly less [36]. While such approaches still require further refinement to enable a resolving power capable of imaging single action potentials, these methods have been able to experimentally confirm that the source of light phase shifts are due to potential-mediated changes in membrane tension, as opposed to swelling due to water transport or electrostriction of the cell membrane [36].

Regarding probing the mechanical response of mammalian cells to electrical excitation, AFM is the most commonly used tool for quantifying cellular deformation despite its invasiveness. Recently, piezoelectric nanoribbons have been developed for electro-mechanical biosensing and have demonstrated that cells deflect by 1 nm when 120 mV is applied to the membrane [37]. Furthermore, these nanoribbons support the model of voltage induced membrane tension discussed earlier, and support previous investigation of cellular electro-mechanics using AFM. Nanoribbons are made using microfabrication techniques, and so can be scaled more readily than AFM probes. Additionally, advances in microfabrication techniques could allow the manufacture of thinner nanoribbons to enhance their sensitivity, and facilitate the electro-mechanical observation of smaller neural structures, such as axons, dendrites, and dendritic spines. The importance of observing the mechanical response of these structures is highlighted further below.

## Mechanical influence imparted by cytoskeletal and extracellular matrices

Mechanobiology is a rapidly growing field investigating the role of mechanical forces in cellular biology and physiology [3]. One common approach to mechanobiology involves the application of structural analysis to the cytoskeletal and extracellular matrices of a cell and determining its associated effects on cellular and molecular processes. Such an approach is based on the concept of tensegrity architecture as a simple mechanical model of cell structure to relate cell shape, movement, and cytoskeletal mechanics, as well as the cellular response to mechanical forces [38]. Tensegrity has allowed the mathematical formulation of the relation of tensioned and compressed parts between the extracellular and cytoskeletal matrices. A model of the intracellular cytoskeleton as a network of interconnected microfilaments, microtubules, and intermediate filaments has been shown to predict dynamic mechanical properties of cells [39].

Another feature of cells that lends well to cellular-mechanical analysis are the intracellular forces of the structural matrices. Intracellular forces may be generated by the polymerization and depolymerization of cytoskeletal elements, such as actin filaments. Actin forms soft macromolecular networks of entangled and cross-linked fibers to establish part of the cellular cytoskeleton. The polymerization and depolymerization of actin filaments and microtubules generates forces that are important to many cellular processes, such as cell motility [40], to counteract plasma membrane tension and deformation changes during clathrin-mediated endocytosis [41], and to act as a molecular tension sensor regulating numerous aspects of intracellular homeostasis and function [42]. The energy for force generation by actin is provided by chemical potential differences between monomeric G-actin and its subunit incorporation into the filamentous F-actin biopolymer. When actin filaments approach a biological load (for example, a plasma membrane) they generate pushing forces where thermal fluctuations enable the continued incorporation of G-actin monomers into the filament. This process of elongation will continue to occur until the counteracting load forces slow and stall polymerization at the thermodynamic limit [43]. This stalling (maximum) force can be estimated by Equation 7:

$$F_{stall} = \frac{k_B T}{\delta} \ln\left(\frac{c}{c_{crit}}\right) \quad (7)$$

where $k_B$ is Boltzmann's constant, $T$ is the absolute temperature, $\delta$ is the elongation distance for a single G-actin monomer (2.7 nm), $c$ is the concentration of G-actin monomers in solution, and $c_{crit}$ is critical concentration for polymerization (equivalent to $k_{off}/k_{on}$) for elongation at a single filament end [44-46]. The maximum force ($F_{stall}$) a single actin filament can generate has been estimated to be ≈ 9 pN.

In a small bundle of actin filaments, it is thought that $F_{stall}$ of the bundle is equal to the linear sum of forces generated by each fiber [45]. In addition to stalling forces, the tips of actin bundles can experience other forces that can cause filaments to buckle. Following buckling, the actin filaments can continue to elongate through the addition of G-actin. The force required to stall actin filament elongation is independent of length, whereas buckling forces vary as a function of length. The force required for filament buckling with one free end and one clamped end can be expressed by Equation 8:

$$F_{buckle} = \frac{\frac{\pi^2}{4} k_{mod}}{L^2} \quad (8)$$

where $k_{mod}$ is the flexural rigidity of an actin filament (0.06 pN/μm$^2$), and $L$ is filament length [45]. To generate forces of several nN/μm$^2$ [47, 48], actin filaments contact surface loads from a variety of different angles and are continuously undergoing nucleation and branch formation for new filament elongation near the leading edge [45].

Actin is in fact one of the best-recognized cytoskeletal contributors to synaptic function. It has been well established that the cytomechanics of axonal growth cone navigation and branching are largely mediated by actin-generated forces [49]. The actin motor proteins myosin are capable of generating forces sufficient to contract muscle tissue and are known to participate in molecular cargo shuffling during cell motility and growth [50]. Recent measurements of growth cone mechanical properties have shown growth cones have a low elastic modulus (E = 106 ± 21 N/m$^2$) and that considering its retrograde flow actin may generate internal stress in growth cones on the order of 30 pN/μm$^2$ [51]. These results indicate growth cones are a soft and weak force generators rendering them sensitive to the mechanical properties of their environment [51], as similarly described above by $F_{stall}$ in Equation 7.

Interestingly, dorsal root ganglion neuron cones have been shown to generate significantly greater traction forces compared to hippocampal neuron growth cones as determined using traction force microscopy [52]. Moreover these neuronal types exhibited differential cytoskeletal adaption to substrate stiffness [52]. Such differences in cytoskeletal mechanics pose the possibility that different forces generated by actin may serve unique mechanical scripts for synapse formation, maturation, and operation in neurons. Perhaps a mechanical environment, such as the extracellular matrix (ECM) within a given anatomical area can change to optimize the growth dynamics of specific groups of invading axons across distinct stages of development. Any such cellular mechanical matching for tuning patterned synapse formation is certainly a tantalizing concept. Several observations seem to provide evidence for such mechanisms. Quantified with atomic force microscopy (AFM), different layers of the hippocampus have been shown to possess significantly different rigidities in the rodent brain (CA1 *stratum pyramidale* = 0.14 nN/μm$^2$, CA1 *stratum radiatum* = 0.20 nN/μm$^2$, CA3 *stratum pyramidale* = 0.23 nN/μm$^2$, and CA3 *stratum radiatum* = 0.31 nN/μm$^2$; [53]). Observed on substrate rigidities ranging from 0.5 and 7.5 nN/μm$^2$, hippocampal axons increase their length faster on softer substrates [54]. Neurons from embryonic spinal cord develop a five-fold higher neurite branch density on soft substrates (0.05 nN/μm$^2$) compared to more rigid ones (0.55 nN/μm$^2$; [55]). The elasticity and mechanical properties of the brain has been shown to change across different stages of development [56-58]. Finally, it has recently been shown that mechanical tension within axons plays an essential role in the accumulation of proteins at presynaptic terminals; biochemical signaling and recognition of synaptic partners is not sufficient [59]. Presynaptic vesicle clustering at neuromuscular synapses vanished upon severing the axon from the cell body and could be restored by applying tension to the severed end, and further stretching of intact axons could even increase vesicle clustering [59]. Furthermore, rest tensions of approximately 1 nN in axons were restored over approximately 15 minutes when perturbed mechanically, implicating mechanical tension as a modulation signal of vesicle accumulation and synaptic plasticity [59]. Increased axonal tension from the resting state may induce further actin polymerization and increased clustering via mechanical trapping or interactions between F-actin and vesicles [59]. Supported by these aforementioned primary observations, neuroscience should focus efforts on characterizing changes in growth cone traction force, elasticity, or viscosity across different anatomical and cellular regions, levels of activity, and stages of development.

Actin in dendritic spines has been shown to regulate synapse formation and spine growth [60], activity-dependent spine motility [61-63], and plasticity [64-66]. In gelsolin knockout mice, reduced actin depolymerization has been shown to enhance NMDA-mediated and voltage-gated (VG) calcium activity in hippocampal neurons [67]. The contribution of mechanical force changes to any of the above observations is not clearly understood. The viscoelasticity of dendritic spines was found to be critical to their function through AFM elasticity mapping and dynamic indentation methods [68]. Through this mechanical characterization, the activity-dependent structural plasticity, metastability, and congestion in the cytoplasm of spines are all gauged by merely a few physically measurable parameters. The degree of which spines are able to remodel and retain stability is determined in large part by viscosity; where soft, malleable spines have properties likely associated with morphological plasticity for learning, and the properties of rigid, stable spines are likely associated with memory retention [68]. Perhaps the stabilization or destabilization of actomyosin networks produces direct mechanical consequences on synaptic activity by increasing or decreasing plasma membrane tension to coordinate the bending or compression of presynaptic compartments and dendritic spines. Given the dynamic nature of the actin cytoskeleton in the regulation of membrane tension and channel activity as further discussed below, the aforementioned idea seems natural for expanded investigations. Additionally, the contribution of the various other elements composing the cytoskeletal and extracellular matrices besides actin as discussed here can analogously be investigated.

## Mechanically-sensitive ion channels

Mechanical forces acting on cell membranes or through cytoskeletal filaments can be transformed into consequences on membrane bound and cytoskeletal-tethered protein activity. Many ion channels exhibit spring-like structures, rendering their gating kinetics sensitive to mechanical forces. Further, the effects of pressure, tension, stretch, and stress on cell membranes are known to be capable of activating and inactivating a broad range of MSCs. From bacteria to primates, nearly all animal cells express MSCs.

To better understand the factors influencing ion channel activity, consider the simple case of a two state (open and closed) channel using Boltzmann statistics, where open channel probability ($P_O$) can be described by Equation 9 as:

$$P_O = \frac{1}{1+e^{\left(\frac{\Delta G}{k_b T}\right)}} \quad (9)$$

where $k_b$ is the Boltzmann constant, $T$ is the absolute temperature, and $\Delta G$ is an intrinsic energy difference between open and closed states ($G_{open} - G_{closed}$). The $\Delta G$ dictates the likelihood of the channel occupying each state. The change in free energy can also be expressed as the sum of changes due to chemical ($\Delta G_{chem}$), electrical ($\Delta G_{elec}$), and mechanical ($\Delta G_{mech}$) contributions. Various analytical models quantifying changes in free energy due to these different mechanisms have been detailed in literature [69-75].

Regarding mechanical stimulation, if a force $f$ is exerted on the channel and the gating domain moves a distance $b$, then work is done, and the change in free energy is described by Equation 10:

$$\Delta G_{mech\ force} = -fb + \Delta u \quad (10)$$

where $\Delta u$ is the intrinsic energy difference between states in the absence of applied force. A larger movement of the gate swing ($b$) requires less force ($f$) to obtain the same $\Delta G$. Note that this expression is equivalent to the gating of voltage-dependent channels as described by Equation 11:

$$\Delta G_{elec\ force} = f_{elec} \times distance = -Eq \times b \approx V_m \left(\frac{qb}{m}\right) \quad (11)$$

where $E$ is the electric field strength and $q$ the net charge on the gating region. The electric field strength is $E = V_m/m$ where $V_m$ is the transmembrane potential and $m$ is the distance over which the field drops, typically approximated as the membrane thickness. Due to difficulties in determining $b$ and $m$, the term $qb/m$ is utilized, and referred to as the equivalent gating charge. The equivalent gating charge is typically 4-6 electron charges per subunit for voltage-gated channels, such that an energy difference of 1 $k_b T$ is produced by a membrane potential of ~5 mV [69].

A similar expression for the free energy holds for channels influenced by tension in the membrane. Tension is the energy excess per unit area resulting from any type of stress. Most work on mechanically gated channels uses patch-recording electrodes to apply suction to a patch of membrane to generate tension. A lateral stretch of a membrane generates an expanded area, producing work, lowering the free energy difference in states, and can be expressed by Equation 12 as:

$$\Delta G_{mech\ tension} = -\gamma \Delta a + \Delta u \quad (12)$$

where $\gamma$ is lateral tension, $\Delta a$ is the change of the in-plane area of the channel after opening, and $\Delta u$ is the intrinsic energy difference between states in the absence of tension. Other forms of stress relevant to biological membranes include shear stress and bending stress, which can also contribute to changes in free energy. However, as the area elasticity modulus is much larger than the shear and bending moduli, the contributions to free energy from lateral tension will typically dominate.

The mechanosensitivity of voltage-dependent channels has been investigated for a multitude of ion channels. Mammalian cells express several families of polymodal-gated ion channels, including transient receptor potential (TRP) and potassium two-pore domain (K$_{2P}$) channels, which have been shown to be activated by mechanical stimuli including membrane stretch and hydrostatic pressure (for reviews see, [76-79]). The TRP channel heteromers TRPC1/C3 and TRPC1/P2 are of particular interest as they have been shown to form calcium-permeable MSCs in mammalian cells including neurons (for reviews see [78-80]). The polymodal K$_{2P}$2.1 channel TREK-1 is mechanosensitive and is active at rest while mediating potassium leak currents to regulate the membrane potential and excitability of neurons [76, 81]. Neurons are also known to express a variety of pressure-sensitive channels, which both exhibit polymodal gating mechanisms. Several of the voltage-gated ion channels expressed in neurons (for example, Na$_V$1.2, Na$_V$1.5, and K$_V$1.1) possess mechanosensitive properties that render their gating kinetics sensitive to transient changes in lipid bilayer tension [69, 77]. In one particular investigation, voltage-gated potassium (K$_V$) channels were shown to exhibit sensitivity to small physiologically relevant mechanical perturbations of the cell membrane, producing shifts in the channel activation curve and an increase in the maximum open probability [82]. Tension sensitivity was accounted for in theory by having the tension act predominantly on the pore opening transition to favor the open conformation, and did not hold if membrane tension acted mainly on the voltage sensor conformational changes [82]. This mechanically-induced shift in activation kinetics could allow K$_V$ channels, and likely other voltage-dependent ion channels related in function and structure such as VG sodium and calcium channels, to play a role in mechanosensation and contribute to the variability of cellular responses to mechanical forces.

The heart is of particular interest for investigating the role of mechanosensitivity in voltage-gated channels, as the heart could represent a more readily accessible model of mechano-electric feedback than the nervous system. If the varying mechanical environment of the myocardium and vasculature are indeed involved in the control of cardiac rhythmicity by the routine deformations of the bilayer membrane structure, then the heart makes for a well-characterized model to study the contributions of stretch-activated currents due to mechano-electric feedback. The heart could also serve as a possible model for investigating the role of mechanically abnormal bilayers (such as in diseased heart) to electrical pathologies of the heart [8]. One early attempt to model the effects of MSCs on heart function include the incorporation of stretch-activated currents into an existing guinea pig ventricular cell model as a linear current introduced by Equation 13 [9]:

$$I_{SAC} = \frac{|V - V_{rev}|\gamma\rho A}{1 + Ke^{-\alpha(L-L_0)}} \quad (13)$$

where the stretch activated current $I_{SAC}$ is determined by the membrane potential $V$, the channel's reversal potential $V_{rev}$, channel conductance $\gamma$, channel density $\rho$, cell area $A$, an equilibrium constant $K$ controlling the amount of current at $L_0$, sarcomere length $L$, and a sensitivity parameter $\alpha$. By using sarcomere length as an analoge of membrane tension, simulated ventricular action potentials successfully captured a number of experimentally observed behaviors. In this manner, more macroscopic characterizations of the cell and tissue can be considered to investigate their effect on stretch-activated currents, and effectively MSCs as well. By examining the stretch-activated currents in heart cells as a whole, in addition to the investigation of the mechanical sensitivity of individual ion channels, further insights on the influence of mechanical forces on the activity of ion channels can be obtained. We should strive to apply these observations and models to nervous systems since it is unimaginable how they might escape factors regulating the influence of mechano-electric coupling on channel activity and cellular excitability.

## Phospholipid membranes

The properties of the cellular phospholipid bilayer membrane determine in part the behavior of various dynamic and relaxation processes. Processes influenced include the propagation and attenuation of mechanical waves, the decay of thermal shape fluctuations, and the translational and rotational diffusion of membrane components [83]. When subjected to lateral stretching or compression, bilayer membranes behave as a viscoelastic material with anisotropy, and this can serve as a mechanism to modulate the state of the membrane, and hence all associated membrane processes such as channel activity. Thermodynamic investigations of lipid phase transitions have shown that lipid density pulses (sound or mechanical waves) can be adiabatically propagated through lipid monolayers, lipid bilayers, and neuronal membranes to influence fluidity and membrane excitability [84-86]. Interestingly, recent evidence indicates such sound wave propagation in pure lipid membranes can produce depolarizing potentials ranging from 1 to 50 mV with negligible heat generation [84], linking mechanical waves in the membrane to changes in transmembrane potentials.

How the properties and changes in density of phospholipid bilayers influence the propagation of mechanical waves and neuronal processes, such as action potential initiation and propagation, is not precisely known. Anesthetics though, make for an interesting case to examine the influence of phospholipid bilayer state on cellular function. It is known that anesthetics affect various functions of the cell, including membrane permeability, hemolysis, nerve function, and the function of ion channels and proteins. The Meyer-Overton rule for anesthetics relates that the critical dose is linearly proportional to the membrane solubility of the anesthetic molecules in the cell membrane, independent of the chemical ligand actions of the molecule [87]. This rules out specific binding effects based on protein models for the wide variety of anesthetics that follow the Meyer-Overton rule. For example, voltage-gated sodium and potassium channels are slightly inhibited by halogenated alkanes and ethers, but not by xenon and nitrous oxide, despite all these anesthetics following the Meyer-Overton rule [87]. It is known however that anesthetics have a pronounced effect on the physical properties of lipid bilayers, such as their lipid melting point phase transitions. This change in physical properties of the membrane can be related to the alteration of cell function by anesthetics, providing a mechanism for the alterations in cell function dependent on their solubility in the membrane and independent of their chemical nature. Such changes in the properties of the cell's membrane would then influence the previously discussed mechanisms of flexoelectricity, voltage induced membrane tension, the forces in cytoskeletal and extracellular matrices, ion channels, and all other mechanically-sensitive processes coupled to the membrane.

## The soliton model of action potentials

An alternate representation of the action potential recently proposed is that of a self-propagating density pulse (soliton) in a cylindrical membrane. This model is better able to explain the phenomena of fluctuations in nerve fiber thickness and reversible heat change associated with the action potential, rather than solely the electro-chemical behavior captured by the HH model [18]. The soliton model of action potentials is based on the thermodynamics and phase behavior of the lipid components of the biological membrane, which are in a fluid state at physiological temperatures. It has been shown that properties of the lipid membrane slightly above melting transition temperature are sufficient to allow the propagation of mechanical solitons in a cylindrical membrane [86].

The soliton model is derived from the wave equation for sound given by Equation 14:

$$\frac{\partial^2 \Delta\rho}{\partial t^2} = \frac{c^2 \partial^2 \Delta\rho}{\partial x^2} \quad (14)$$

where $\rho$ is the lateral density of the nerve membrane, $\Delta\rho$ is the change in density compared to the resting membrane, $t$ is time, and $x$ is position. The propagation velocity $c$ is typically considered constant when describing sound propagation in air, however, the speed of sound is a sensitive function of density and frequency close to the melting transition in membranes [88]. This nonlinearity of both the density and frequency dependence of sound velocity makes soliton propagation possible. The differential equation resulting from expansion of the wave equation is complex, but can be expressed as the following localized analytical solution [87] describing the shape of a propagating density excitation through Equation 15:

$$\Delta\rho^A(z) = \frac{p}{q} \cdot \frac{1 - \left(\frac{v^2 - v_{min}^2}{c_0^2 - v_{min}^2}\right)}{1 + \left(1 + 2\sqrt{\frac{v^2 - v_{min}^2}{c_0^2 - v_{min}^2}} \cosh\left(\frac{c_0}{h} z \sqrt{1 - \frac{v^2}{c_0^2}}\right)\right)} \quad (15)$$

where $\Delta\rho^A$ is the area lateral density change, $z$ is the position along the axon, the parameters $p$ and $q$ describe the dependence of the sound velocity on density, $v$ is the propagation velocity, $v_{min}$ is the minimum velocity allowed for soliton propagation, $c_0$ is the velocity of small amplitude sound, and $h$ is a parameter describing the frequency dependence of the speed of sound (dispersion). As required by the equation, and observed experimentally [86], the soliton profile has a maximum $[\partial(\Delta\rho^A)/\partial z = 0]$ about which it is symmetric.

Of particular interest is the comparison of the energy carried by solitons with the electrostatic energy associated with the conventional modeling of pulse propagation in a nerve. If the empirically observed energy is greater than the electrostatic energy, then the conventional HH mechanism for pulse propagation is insufficient. This inequality has in fact been shown, as the electrical energy released and reabsorbed from a membrane capacitance is unable to account for more than half of observed temperature changes during an action potential [89]. The associated energy of a propagating soliton will have potential and kinetic energy contributions, which, using a Lagrangian formalism, the energy density is given by Equation 16 [86]:

$$e_{sol} = \frac{c_0^2}{\rho_0^A} (\Delta\rho^A)^2 + \frac{p}{3\rho_0^A} (\Delta\rho^A)^3 + \frac{q}{6\rho_0^A} (\Delta\rho^A)^4 \quad (16)$$

Assuming a maximum voltage change at the peak of the soliton, $V_0$, and that the capacitive energy of a membrane is due to compression and the accompanying voltage change, the capacitive energy density is given by Equation 17 [86]:

$$e_{cap} = \frac{1}{2} C \cdot \left(\frac{V_0 \Delta\rho^A}{\Delta\rho_{max}^A}\right) \quad (17)$$

where $\Delta\rho_{max}^A$ is the maximum amplitude of the lateral density change of the membrane and $C$ is the capacitance of the membrane. Comparisons of the estimates of energy using these equations found that the electrostatic energy density was more than one order of magnitude smaller than the energy of the corresponding soliton [86].

Clearly, the HH model alone cannot capture a number of behaviors of the propagating nerve pulse, such as the reversible transfer of heat and mechanical changes that are captured in the soliton model of propagating activity. However, proteins do not function as active components or channels in the model of soliton propagation through the phospholipid bilayer, but rather tune the thermodynamics of the membrane. Whether initiation of the mechanical soliton is used for communication is therefore left unanswered, and the numerous investigations supporting the role of proteins in the electrical propagation of nervous signals are left neglected. Also, the collision of action potentials is known to block the propagation of nervous signals. The collision of solitons according to the above equations based on adiabatic and reversible physics though, allows pulses to pass through each other with minimal loss of energy [90]. The inclusion of proteins for electro-mechanical coupling may resolve these issues, however the combination of the soliton and HH models, which are based in separate mechanisms, is not straightforward.

Hypothesis of neuro-mechanical signaling

Beyond the role of anesthetics to change the mechanical properties of the lipid bilayer, and hence cell function, it has recently been proposed that propagating density pulses in the membrane may serve as the actual signal that modulates the function of membrane bound enzymes, operating as an alternative mechanism for cellular signaling and nerve pulse propagation [91-93]. In this hypothesis, the activity or conformational fluctuations of one protein initiates a local mechanical disturbance, which propagates along the lipid interface to another protein, transiently changing the thermodynamic state of the second protein's environment and influencing its activity. Communication of this form would not require energy since mechanical pulse propagation is adiabatic and enzymes work reversibly. Additionally, the dielectrical properties of the interface would lead to a propagating voltage pulse coupled to these density oscillations as discussed earlier.

Dynamic studies of the biological lipid interface using thermodynamic concepts, rather than the currently prevailing electrical theory, were used to explore this hypothesis. By monitoring 2D pressure pulses in lipid monolayers, the degree of excitability of the interface was found to depend on its thermodynamic state. Close to the maximum of compressibility of the interface, the pulse signal becomes very weak, illustrating that the thermodynamic state of the interface influences propagation speed and strength [92]. Furthermore, by investigating different interfaces at different temperatures, correlations of the optical and mechanical states of lipid monolayers was found to be a property of the thermodynamic state, and not due to the nature of the molecules investigated [93]. This coupling between fluorescent intensity and pressure pulses were clearly resolved when the system was excited within or nearby the transition region of the membrane, as was the condition for self-propagating solitons discussed earlier.

To investigate the phenomena in biological systems, the temperature dependence of an excitable medium's mechanical material properties were derived and also correlated with temperature-dependent relaxation of pulse propagation in blackworms, nerves, and gels [91]. It is typically assumed that temperature sensitivity of the duration of refractory period is determined by the timescales required by metabolic processes to reestablish resting ion gradients and channel kinetics. However, using solely the framework of thermodynamic theory and no assumptions about metabolic reactions or equilibration processes, the predictions of relaxation times based solely on theory compared well with experimental data. The velocity-temperature relationship for vessel pulsations followed a conserved pattern for pulses in excitable systems: linear increase of velocity with increasing temperature ultimately interrupted by a heat block. It was concluded that conservation of temperature dependence of pulse propagation velocity is likely a consequence of some well conserved physical mechanism, such as mechanical state, and not dependent on metabolic reactions [91]. Thus, oscillations in the thermodynamic state of the lipid interface is sufficient to propagate density pulses, and as required thermodynamically, any proteins or enzymes located in the path of such a density pulse will exhibit an altered kinetic behavior. These observations and conclusions form the basis of the proposed hypothesis of mechanical signaling along the cell membrane.

## Mechanically interfacing to nervous systems

The modulation and monitoring of the nervous system is pertinent to the treatment of neurologic and psychiatric diseases, as well as the scientific investigation of the neural mechanisms of cognitive, sensory, and motor functions. Conventionally, interfacing with the nervous system has been conducted using electrical and chemical means, such as micro-dialysis and deep-brain stimulation. Recently, the development of devices utilizing mechanical energy to interact with the nervous system has received considerable attention. These devices include ultrasound for noninvasive neural stimulation [94] and magnetic resonance elastography for noninvasive palpitation and mechanical characterization of the brain [95].

Besides its use for diagnostic imaging, ultrasound at low intensity is able to nondestructively excite nervous tissue [94, 96]. The mechanisms behind ultrasound stimulation however are not well established. There are two classes of mechanisms primarily considered, thermal and mechanical. Ultrasound can heat tissue, analogous to transcranial high-intensity focused ultrasound ablation [97-99], and temperature sensitive ion channels can be activated through tissue heating. However, negligible temperature increases have been measured during pulsed ultrasound stimulation protocols [94, 100-102]. The premise behind the hypothesized mechanical mechanisms of ultrasound is that deformation of the cell membrane, or the proteins embedded therein, could affect ion channel kinetics and/or membrane capacitance to induce transmembrane currents to initiate action potential discharge [103, 104]. Recently, intramembrane cavitation has been proposed as a mechanism for the effects on nervous tissue by ultrasound [105, 106]. Using models of the cellular membrane, the mechanical energy from ultrasound would be absorbed and transformed by the membrane into expansions and contractions of the space between bilayer membrane leaflets [105]. Linking this model with electro-deformation, ultrasound led to action potential excitation via currents induced by membrane capacitance changes within the computational model [106]. This model is referred to as the bilayer sonophore, and offers explanations on the requirement for long ultrasonic stimulation pulses and other experimentally observed phenomena.

Magnetic resonance imaging (MRI) is the most common imaging modality for investigating central neurological disorders as it is noninvasive and provides a number of contrast mechanisms. Magnetic resonance elastography (MRE) determines the shear modulus of tissues *in vivo* through the application of mechanical shear waves and the use of a phase sensitive magnetic resonance imaging sequence to produce a map (elastogram) of the shear modulus of the tissue. The elastogram is used for clinical diagnostics, as a change in cellular elasticity is associated with many diseases [107], due to their altering the microstructural environment of the central nervous system through neuroinflammation, neurodegeneration, and disruption of the glial matrix. This is analogous to the palpitation of tissue to identify lesions based on their differential stiffness to surrounding tissues such as breast tumors. Application of MRE to the brain may have useful applications for characterizing brain disease based on the mechanical properties of the tissue. The mechanical properties of the human brain have shown a high sensitivity to neurodegeneration in initial investigations of Alzheimer's disease, multiple sclerosis, normal-pressure hydrocephalus, and cancer [108]. However, MRE has yet to gain traction in clinical applications as mechanical properties are largely reported as global averages rather than local values. Recent work on MRE has been focused on generating high-resolution, reliable, and repeatable estimates of local mechanical properties in the human brain. The mechanical properties of the corpus callosum and corona radiate were recently measured in healthy individuals using high-resolution MRE and atlas-based segmentation [109]. Both structures were found to be stiffer than the overall white matter, and demonstrated the feasibility of quantifying the mechanical properties of specific structures in white matter architecture for the assessment of the localized effects of disease. The ability to reliably estimate local mechanical properties noninvasively represents a possible revolution in the clinical assessment of neurodegeneration of the human brain.

## Conclusions

As discussed throughout this quantitative perspective, the HH model alone does not capture a number of biophysical phenomena associated with propagating nerve impulses or action potentials. With respect to physical forces, the significance of dynamic mechanical changes occurring in membranes during action potential propagation remains poorly understood [1, 2]. However, the mechanisms underlying these associated processes of the action potential have been investigated and modeled separately. Based on these observations we should consider the importance of electro-mechanical coupling on neuronal excitability since modulation of membrane tension by the transmembrane voltage of a neuron causes movement of the membrane with magnitude and polarity governed by the cell membrane stiffness and surface potentials in order to maintain pressure across the membrane.

Mechanical forces acting on cell membranes or through cytoskeletal filaments can also be transformed into consequences on membrane bound and cytoskeletal tethered protein activity. Membrane bound protein activity is also influenced by the properties of the cellular phospholipid bilayer. How the properties and changes in density of phospholipid bilayers influence the propagation of mechanical waves and neuronal processes, such as action potential initiation and propagation, is not precisely known. It has recently been proposed that propagating density pulses in the membrane may also serve as the signal that modulates the function of membrane bound enzymes, operating as an alternative mechanism for cellular signaling and nerve pulse propagation. The development of devices utilizing mechanical energy to interact with the nervous system has received considerable attention recently, and includes ultrasound for noninvasive neural stimulation and magnetic resonance elastography for noninvasive palpitation of the brain.

The extent to which cellular-mechanical dynamics influences neuronal activity, and effectually the interfacing to the nervous system using mechanical forces, remains largely unexplored. To advance neuroscience and our understanding of the complex nervous system, the compartmentalization of analyses and processes due to electrical, chemical, or mechanical energies in system characterization and manipulation needs to be stepped away from. While numerous mechanical events have been observed and associated with neuronal activity, it has not been until very recently that technology has started to be adapted to capitalize on these mechanical events to allow the observation, and even modulation, of nervous tissue. By starting to consider the interplay between electrical, chemical, thermal and mechanical energy, rather than separately compartmentalizing them, fresh insights into nervous system function and dysfunction will likely evolve. We are certainly curious to learn how neuroscience and physics will think about neural function 20 years from now.

## Acknowledgements

We graciously thank the supporters of our research and development efforts. Our work is supported by grants to WJT from The McKnight Endowment Fund for Neuroscience, Neurotrek, Inc., and faculty start-up funds provided by the Virginia Tech Carilion Research Institute. JM is a doctoral student in the Virginia Tech – Wake Forest School of Biomedical Engineering and Sciences and is supported by a Virginia Tech Carilion Medical Research Scholar Fellowship. For further information regarding our research efforts, visit: http://www.tylerlab.com.

## Disclosure

WJT is a co-founder, Chief Science Officer, and Board of Directors Member of Neurotrek, Inc. He is also an inventor on 20+ patent applications describing systems, methods, and devices for non-invasive brain circuit modulation. JM is also an inventor on a patent application describing brain modulation systems, methods, and devices.